# THE NEED TO MEASURE LOW ENERGY, ANTI-NEUTRINOS ($E_\nu$ <0.782 MeV) FROM THE SUN

©2002 O. Manuel

University of Missouri, Nuclear Chemistry, Rolla, MO 65401 USA

Received (             )

Measurements are needed of low energy, anti-neutrinos generated by possible neutron decay at the core of the Sun. The measurement will test the validity of a proposal that solar luminosity, solar neutrinos, and the outpouring of $H^+$ ions from the solar surface are products of a chain of reactions triggered by neutron emission from the solar core. Inverse β-decay of 87 day $^{35}S$, induced by capture of low-energy anti-neutrinos on $^{35}Cl$, is a likely candidate for this measurement.

Fresh debris of a supernova (SN) formed the Solar System. This is the conclusion to numerous measurements since 1960 [1]. The SN exploded about 5 Gy ago [2]. The Sun formed on the collapsed SN core, Fe-rich material around the SN core formed iron meteorites and cores of the inner planets, and material from the outer SN layers formed the giant Jovian planets from light elements like H, He and C [3,4]. This sequence of events is shown in Fig. 1.

When first proposed in the mid-1970's [3,4], the scene in Fig. 1 appeared to contradict several widely held opinions:

1. Supernovae always explode symmetrically.
2. SN debris cannot form planets surrounding the collapsed SN core.
3. Poorly mixed SN debris is inconsistent with uniform atomic weights and isotopic abundances assumed throughout the solar system.
4. The Sun is mostly hydrogen; it could not be formed in this manner.
5. Abundant solar hydrogen is required to explain solar luminosity.

This paper suggests a measurement to test the validity of the 5th assumption. The other objections have been resolved.

Problems with the 1st assumption were exposed when the Hubble telescope found that asymmetric SN explosions are commonplace, for example, the explosion of SN1987A [5]. The validity of the 2nd assumption was brought into question with the finding of two rocky, Earth-like planets and one Moon-like object orbiting a neutron star, PSR1257+12 [6]. Decay products of nuclides even shorter-lived than $^{129}I$ and $^{244}Pu$ [7,8] and linked elemental and isotopic variations in meteorites and planets [9-14] ruled out the 3rd assumption.

The 4th assumption, that the interior of the Sun is hydrogen-filled like its surface, was invalidated by the finding that mass separation inside the Sun enriches lighter elements and the lighter isotopes of each element at the solar surface [15]. After correcting for mass fractionation, the most abundant elements in the interior of the Sun [15] were shown to be the same elements Harkins [16] found to comprise 99% of ordinary meteorites, i.e., Fe, O, Ni, Si, Mg, S, and Ca. These elements are made in the deep interior of highly evolved stars [17], as expected from the events in Fig. 1.

Supernova debris is not the only possible source for some short-lived nuclides and isotopic anomalies observed in meteorites and planets [9-14]. However, rapid neutron capture in a supernova is the only viable mechanism to produce extinct $^{244}$Pu [2,8]. The validity of Fig. 1 was strengthened when Kuroda and Myers [2] combined age dating techniques based on $^{238}$U, $^{235}$U and $^{244}$Pu to show that a supernova explosion produced these actinide nuclides at the birth of the Solar System, about 5 Gy ago. Their results are in Fig. 2.

The present paper concerns the need for measurements to test a proposed source for luminosity in an iron-rich Sun, i.e., to test the validity of the 5th objection to events shown in Fig. 1. In layman's terms, "How can the Sun shine if made mostly of iron?" The answer requires the use of reduced variables, like those in van der Waals' equation of corresponding states, to look for sources of energy in properties of the 2,850 nuclides tabulated in the latest report from the National Nuclear Data Center [18].

The results are shown in Fig. 3, where data for ground states of 2,850 known nuclides [18] are displayed on a 3-dimensional graph of Z/A, charge per nucleon, versus M/A, mass or total potential energy per nucleon, versus A, mass number. All nuclides lie within the limits of 0≤Z/A≤1 and 0.998≤M/A≤1.010, and the data [18] define a cradle shaped like a trough or valley. The more stable nuclides lie along the valley, and the lowest point is $^{56}$Fe.

Cross-sections through Fig. 3 at any fixed value of A reveal the familiar "mass parabola". An example is shown in Fig. 4 at A = 27. Data points are also shown in Fig. 4 for unbound nucleons, $^1$n on the left at Z/A = 0 and $^1$H on the right at Z/A = 1.0. The mass parabola in Fig. 4 is defined by masses of ground state nuclides at $^{27}$F, $^{27}$Ne, $^{27}$Na, $^{27}$Mg, $^{27}$Al, $^{27}$Si, $^{27}$P and $^{27}$S [18]. At Z/A = 0, the empirical parabola yields a value of M/A = M($^1$n) + ~10 MeV.

The results shown in Fig. 4 are typical of values indicated at Z/A = 0 from mass parabolas at each mass number A, where A>1. Together these mass parabolas [18] at all values of A suggest a driving force of ~10-22 MeV for neutron-emission from a neutron star [19]. This energy source and the following sequence of reactions thus offer a possible explanation for solar luminosity (SL), the 5th objection to the scene shown in Fig. 1:

- Neutron emission from the solar core (>57% SE)
    - $<^1n> \rightarrow\ ^1n\ +\ \sim 10\text{-}22$ MeV
- Neutron decay or capture (<5% SE)
    - $^1n \rightarrow\ ^1H + e\text{-} + \text{anti-}\nu\ + 0.782$ MeV
- Fusion and upward migration of H+ (<38% SE)
    - $4\ ^1H+ + 2\ e\text{-} \rightarrow 4He^+ + + 2\ \nu\ + 27$ MeV
- Escape of excess H+ in the solar wind (100% SW)
    - Each year $3 \times 10^{43}$ $H^+$ depart in the solar wind

Thus neutron-emission from a collapsed SN core in the center of the Sun, as shown in Fig. 1, may start a chain of reactions that explain luminosity, as well as the observed outflow of solar neutrinos and H+ ions from the surface of an iron-rich Sun [19]. In the first step, neutron emission may release 1.1% - 2.4% of the nuclear rest mass as energy. In the third step, hydrogen fusion releases about 0.7% of the rest mass as energy.

However, this paper is most concerned with the need to measure low energy, anti-neutrinos emitted in the second step. This measurement may confirm or deny the occurrence of the above reactions in the Sun and the historical validity of the events shown in Fig. 1.

Specifically, it is proposed to look for inverse β-decay induced by low-energy anti-neutrinos coming from the Sun, e.g., $^3He \rightarrow\ ^3H$; $^{14}N \rightarrow\ ^{14}C$; or $^{35}Cl \rightarrow\ ^{35}S$. The latter reaction in the Homestake Mine [20] might produce measurable levels of 87-day 35S. Alternatively, $^{35}S$ might be extracted from underground deposits of salt (NaCl) and detected by counting or by AMS (Accelerator Mass Spectrometry) measurements.

This work was supported by the University of Missouri and by the Foundation for Chemical Research, Inc.

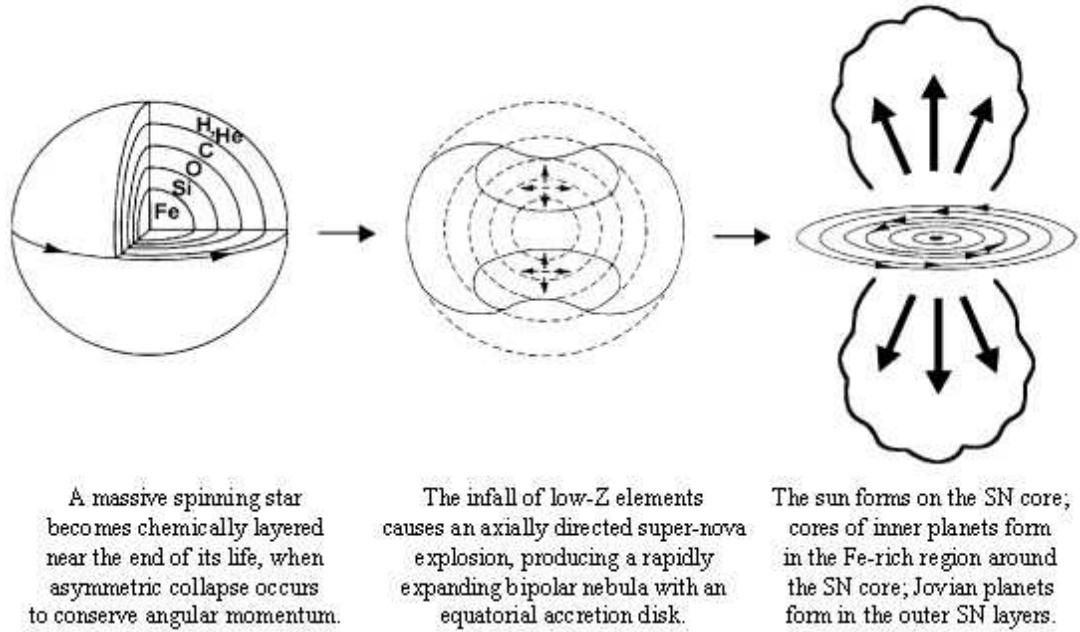

Fig. 1. Formation of the Solar System from the chemically and isotopically heterogeneous debris of a spinning supernova [3,4]. The Sun formed on the collapsed supernova core, Fe-rich material formed iron meteorites and cores of the inner planets, and H-, He- and C-rich material from the outer layers formed giant planets like Saturn and Jupiter.

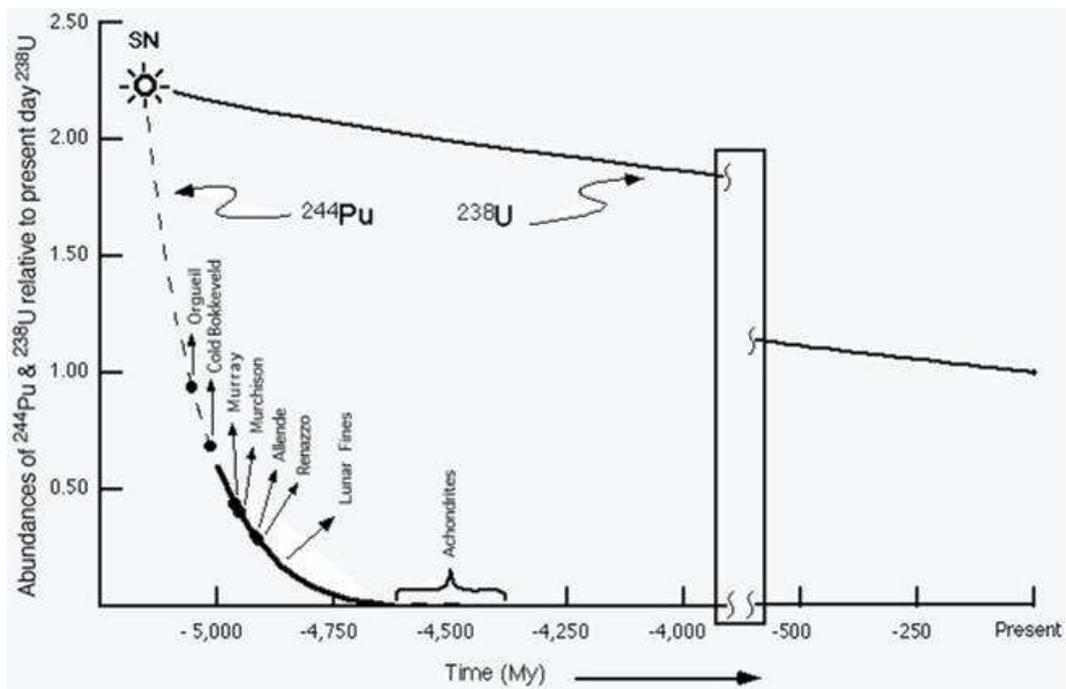

Fig. 2. Combined U-Pb and Pu-Xe age dating of the supernova explosion that occurred at the birth of the Solar System, about 5 Gy ago [2].

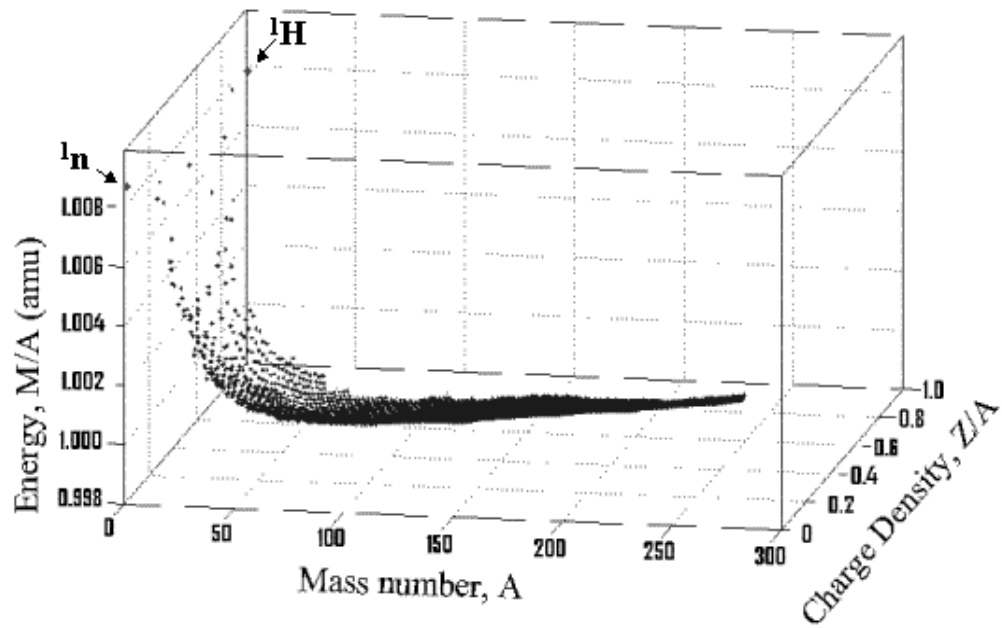

Fig. 3. The "Cradle of the Nuclides", a 3-dimensional plot of reduced nuclear variables, Z/A (charge per nucleon) versus M/A (mass per nucleon) versus A (mass number) for the ground states of 2,850 known nuclides [18]. The cradle provides new insight into possible sources of nuclear energy [19].

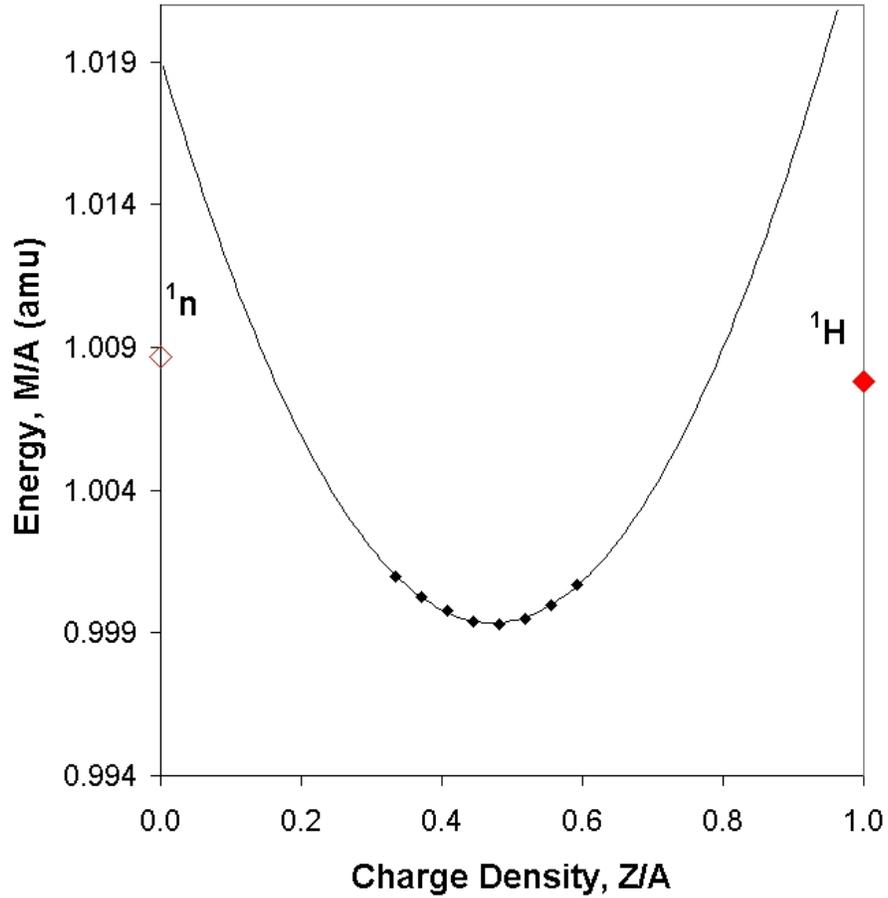

Fig. 4. A cross-sectional cut through the "Cradle of the Nuclides" (Fig. 3) at A = 27. The mass parabola is defined by masses of ground state nuclides at 27F, 27Ne, 27Na, 27Mg, 27Al, 27Si, 27P and 27S. For comparison, reference values are shown for unbound nucleons, 1n on the left at Z/A = 0 and 1H on the right at Z/A = 1.0.